\documentclass[aps,prb,twocolumn,floatfix,superscriptaddress]{revtex4-2}
\usepackage{graphicx}
\usepackage[utf8]{inputenc}
\usepackage{amssymb}
\usepackage{amsmath}
\usepackage[colorlinks,allcolors=blue]{hyperref}
\usepackage{braket}
\usepackage{placeins}
\usepackage{soul}
\usepackage{etoolbox}

\begin{document}

\setlength\abovedisplayskip{5pt}
\setlength\belowdisplayskip{5pt}

\title{Thermalization in the mixed-field Ising model: An occupation number perspective}

\author{Isa\'ias Vallejo-Fabila}
\affiliation{Department of Physics, University of Connecticut, Storrs, Connecticut 06269, USA}
\author{Fausto Borgonovi}
\affiliation{Dipartimento di Matematica e
  Fisica and Interdisciplinary Laboratories for Advanced Materials Physics,
  Universit\`a Cattolica, via della Garzetta  48, 25133 Brescia, Italy}
\affiliation{Istituto Nazionale di Fisica Nucleare,  Sezione di Milano,
  via Celoria 16, I-20133,  Milano, Italy}
\author{Felix M. Izrailev}
\affiliation{Instituto de F\'{i}sica, Benem\'{e}rita Universidad Aut\'{o}noma
  de Puebla, Apartado Postal J-48, Puebla 72570, Mexico}
\affiliation{Department of Physics and Astronomy, Michigan State University, E. Lansing, Michigan 48824-1321, USA}
\author{Lea F. Santos}
\affiliation{Department of Physics, University of Connecticut, Storrs, Connecticut 06269, USA}

\begin{abstract}
The occupation number is a key observable for diagnosing thermalization, as it connects directly to standard statistical laws such as Fermi--Dirac, Bose--Einstein, and Boltzmann distributions. In the context of spin systems, it represents the population of the sublevels of the magnetization in the $z$-direction. We use this quantity to probe the onset of thermalization in the isolated quantum and classical one-dimensional spin-1 Ising model with transverse and longitudinal fields. Thermalization is achieved when the long-time average of the occupation number converges to the microcanonical prediction as the chain length $L$ increases, consistent with the emergence of ergodicity. However, the finite-size scaling analysis in the quantum model is challenged by the exponential growth of the Hilbert space with $L$. To overcome this limitation, we turn to the corresponding classical model, which enables access to much larger system sizes. By tracking the dynamics of individual spins on their three-dimensional Bloch spheres and employing tools from random matrix theory, we establish a quantitative criterion for classical ergodicity in interacting spin systems. We find that deviations from classical ergodicity decay algebraically with system size. This power-law scaling then provides a quantitative bound on the approach to thermal equilibrium in the quantum model. 
\end{abstract}

\maketitle

\section{Introduction}
\label{sec:INT}
Thermalization in classical Hamiltonian systems relies on mixing, where trajectories lose memory of initial conditions and correlations decay as phase-space distributions evolve toward a uniform distribution on the energy shell. Because mixing implies ergodicity, it ensures that time averages coincide with ensemble averages, thereby providing the dynamical foundation of equilibrium statistical mechanics~\cite{KrylovBook}. Chaos is the dynamical mechanism that generates mixing~\cite{Sinai1970,Chirikov1979,TaborBook,GutzwillerBook,OttBook}: it is characterized by exponential sensitivity to initial conditions quantified by positive Lyapunov exponents. However, instability alone is not sufficient for chaos, as in the case of the inverted pendulum near its unstable fixed point. Likewise, ergodicity does not necessarily imply chaos -- e.g. a particle moving on a rectangular billiard with irrational velocity components explores the energy shell ergodically despite exhibiting completely regular dynamics. Thus, thermalization in classical systems requires not only instability or ergodicity alone, but the combination of both. 

When applying the concept of ergodicity to physical systems, one must recognize that the mathematical ergodic theory is formulated in the idealized limit of time $t \to \infty$, which introduces a double-limit problem when studying thermalization, since the long-time limit may not commute with the thermodynamic limit $L \to \infty$, where $L$ is the system size. Moreover, ergodicity cannot be rigorously established through numerical simulations, since finite-time computations cannot exclude the existence of arbitrarily small regular islands in phase space. An example is the kicked rotor on the torus. For large kicking strength, its dynamics appears fully chaotic and numerically indistinguishable from an ergodic motion, yet small stable islands may persist~\cite{Chirikov1979}. Although such regions are negligible from a physical standpoint, they illustrate the distinction between strict mathematical ergodicity and the practical notion of effective ergodicity relevant for physical thermalization.

Similarly to the classical limit, thermalization in isolated quantum systems requires that at long times, the time averages of physical observables coincide with the predictions of statistical mechanics. However, in contrast to classical mechanics, where chaos arises from nonlinear equations of motion leading to exponential sensitivity to initial conditions, quantum dynamics is unitary and governed by the linear Schr\"odinger equation. As a result, the notions of chaos and ergodicity from classical physics do not carry over directly, and their distinction in the quantum domain becomes subtle and often blurred~\cite{Pilatowsky2023,Aravinda2024,Vikram2025}. From a mathematical standpoint, quantum ergodicity is formulated in the semiclassical limit~\cite{Shnirelman1974,Zelditch1987,Shnirelman1993,Shnirelman2023}. In physics, it is often linked to Berry’s conjecture~\cite{Berry1977}, which states that energy eigenfunctions behave like superpositions of random plane waves with uncorrelated phases. This is exemplified by full random matrix ensembles~\cite{MehtaBook,Brody1981}, whose eigenstates can be treated as random states,  fully delocalized over the entire Hilbert space. 

Just as classical ergodicity cannot be rigorously established in physical systems at finite times, quantum ergodicity in the sense of full random-matrix behavior is not achieved in realistic many-body quantum systems, where interactions are few-body and of finite range. Because of this locality, the Hamiltonian matrix in a physically meaningful basis is sparse and interactions couple only a subset of basis states $|k\rangle$, so that each Hamiltonian eigenstate, $|\alpha \rangle = \sum_k C_k^{\alpha} |k \rangle$, spreads over only a portion of the full Hilbert space. In the energy representation, the maximum subset of basis states associated with the eigenstate defines its energy shell. When the number of dominant components $C_k^{\alpha}$ occupies only a small fraction of the shell, the eigenstates are localized. In contrast, when the coefficients behave as random Gaussian variables distributed around the smooth envelope of the energy shell, the eigenstates are effectively  ergodic~\cite{ZelevinskyRep1996,Santos2012PRE,Santos2012PRL,Borgonovi2016}. This restricted yet physically meaningful notion of ergodicity is sufficient to ensure thermalization of few-body observables in isolated quantum systems. It emerges in the chaotic regime~\cite{Flambaum1994}, where spectral correlations follow random matrix theory~\cite{Guhr1998} and the eigenstates exhibit maximal spreading within the energy shell~\cite{Santos2012PRE,Santos2012PRL}.  Such states -- often referred to as chaotic eigenstates -- typically occur close to the middle of the many-body spectrum.   

In this work, we examine thermalization from the perspective of the occupation number. This quantity serves as a sensitive probe of thermalization, because at thermal equilibrium the populations of single-particle (single-quasiparticle) levels become stationary and follow the appropriate Maxwell--Boltzmann, Bose--Einstein, or Fermi--Dirac distribution depending on particle statistics and conserved quantities. Furthermore, the occupation number distribution   determines the equilibrium values of observables that depend on single-particle populations, such as energy and particle number. By analyzing the dynamics and scaling of occupation numbers in both the quantum model and its classical counterpart, we aim to establish a concrete link between the microscopic structure of eigenstates and macroscopic thermodynamic behavior.

The idea of using occupation-number statistics to diagnose thermalization in isolated many-body quantum systems traces back to pioneering works~\cite{Srednicki1994,Horoi1995,Flambaum1997,Borgonovi1998}. In Ref.~\cite{Srednicki1994}, it was shown that if individual many-body eigenstates satisfy Berry’s conjecture, then the momentum-occupation profile computed within a single eigenstate already reproduces the Maxwell--Boltzmann distribution. In studies of interacting systems of fermions and bosons, it was demonstrated that for individual chaotic many-body  eigenstates~\cite{Flambaum1994}, the resulting equilibrium distributions assume the conventional Bose--Einstein~\cite{Horoi1995,Borgonovi1998} and Fermi--Dirac~\cite{Flambaum1997} forms. In this framework, thermodynamic parameters such as temperature emerge intrinsically from the properties of a single eigenstate and can be expressed in terms of energy, particle number, and interaction strength. 

Our analysis focuses on spin models, for which we adapt the concept of occupation numbers to represent the average population per magnetic sublevels. If the spin dynamics are ergodic, 
the long-time distribution of these populations should coincide with the microcanonical prediction, implying that all magnetic sublevels become equally populated.

We consider the one-dimensional (1D) Ising model with both transverse and longitudinal magnetic fields, where  the interplay between interactions and competing fields give rise to classical and quantum chaos. Unlike most existing studies, which focus on the spin-1/2 version of this model~\cite{Monasterio2005,Karthik2007,Banuls2011,Atas2017,Alcantara2019,Noh2021,Bhattacharjee2024,Rodriguez2024}, we investigate spins with quantum number $S=1$. We select parameters that maximize chaotic behavior and initial states near the middle of the spectrum to analyze how the quantum system approaches thermalization in the thermodynamic limit. Even though the difference between the infinite-time and thermal averages of on-site magnetizations and occupation numbers slightly decreases as the system size $L$ increases, the exponential growth of the Hilbert space limits the accessible system sizes, preventing a proper scaling analysis.

To address this issue, we turn to the classical limit~\cite{Wijn2012,Lebel2023,Benet2023,Borgonovi2025,Ermakov2025}, where simulations can be performed for much larger system sizes and the notions of ergodicity and chaos are well defined. Our choice of $S=1$ ensures a meaningful quantum-classical correspondence~\cite{Emerson2001,Elsayed2015,Benet2023,Borgonovi2025}. Importantly, our analysis explicitly compares the quantum and classical models rather than relying on any semiclassical approximation.

As noted above, classical thermalization requires both positive Lyapunov exponent and ergodicity. Using random matrix theory, we provide a rigorous definition of classical ergodicity for spin chains, based on the motion of each spin on its three-dimensional (3D) sphere. Applying this to the 1D chaotic mixed-field Ising model, we find that deviations from ergodicity decay as a power-law of the system size. Assuming that non-ergodicity in the classical limit also implies its absence in the quantum regime, this scaling establishes a quantitative bound on the approach to thermalization in the corresponding quantum system.  

The algebraic convergence toward thermal behavior is consistent with previous findings for average values of few body-observables~\cite{Rigol2008}, but it contrasts with the widely held expectation of exponentially fast convergence between infinite-time and thermal averages in chaotic quantum systems~\cite{Ikeda2011}. While the temporal fluctuations of observables after equilibration indeed decay exponentially with system size~\cite{Zangara2013,Lezama2023,footIsaias}, our results show that the approach to thermalization itself follows a slower, power-law behavior.

The paper is organized as follows. Section~\ref{Sec:Model} analyzes the quantum properties of the system, including level statistics, eigenstate structure, thermalization indicators, and the relaxation of the occupation number. Section~\ref{sec:ClChaos} turns to the classical limit. We first identify the regime of chaotic dynamics through Lyapunov exponents, then establish an analytical criterion for classical ergodicity using random matrix theory, and apply it to the scaling with system size of the physical spin model. The section ends by establishing a correspondence between the classical and quantum behavior. Section~\ref{sec:CON} summarizes our main findings.

\section{Quantum model: Chaos and Thermalization}
\label{Sec:Model}

In this section, we demonstrate that both the level statistics and the structure of the eigenstates of the quantum mixed-field Ising model with spin $S=1$ closely follow the predictions of standard random matrix theory across a broad range of parameters. This correspondence implies that expectation values of local observables, such as the onsite $z$- and $x$-magnetizations, approach their microcanonical values and that their off-diagonal matrix elements exhibit Gaussian statistics, as we explicitly verify. Building on this equilibrium analysis, we then investigate the quench dynamics of the occupation number, comparing its infinite-time average to the microcanonical prediction and exploring how the agreement scales with system size.

The Hamiltonian for the 1D mixed-field Ising model with $L$ sites and open boundary conditions is given by 
\begin{equation}
 H=   -J \sum_{j=1}^{L-1}  S_j^z S_{j+1}^z -g\sum_{j=1}^L S_j^z   - h\sum_{j=1}^L S_j^x, 
 \label{eq:MFIM}
\end{equation}
where $S_j^{\mu}$, with $\mu=x,y,z$, represent spin operators acting on site $j$, $J$ is the Ising interaction strength between neighboring sites, which we fix as $J=1$, and the amplitudes of the uniform fields in the $z$- and $x$-directions are, respectively, $g$ and $h$.  
In many of our studies below, we focus on the central site, which we denote by $j=c= \lceil L/2 \rceil$, where $\lceil x \rceil$ means the smallest integer greater than or equal to $x$. Since we only consider odd values of $L$, this simplifies to $c =(L+1)/2$.

The Hamiltonian conserves parity (reflection symmetry) and is trivially integrable when $h=0$ for any $S$. For spin-1/2 and $g=0$, the model maps onto a system of free fermions via the Jordan-Wigner transformation~\cite{FranchiniBook}, making it exactly solvable, while quantum chaos emerges when all parameters are non-zero ($g,h,J \neq 0$) \cite{Monasterio2005,Banuls2011,Bhattacharjee2024}. In fact, chaos can even be triggered by setting $g\neq0$ on a single site only while keeping $h,J \neq 0$ \cite{Santos2020}.

Instead, we focus on the spin-1 model for which fewer results are available in the literature. In this case, nonintegrability persists even in the absence of a longitudinal field ($g=0$). In the following, we denote the eigenvalues and eigenstates of $H$ by $E_\alpha$ and $|\alpha \rangle$, respectively.

To facilitate comparison with the classical model, we introduce an effective Planck constant,
$$\hbar_{\text{eff}} =1/\sqrt{S(S+1)},$$ 
such that the spin vector satisfies  $|\vec{S_j}^2| = \hbar_{\text{eff}}^2 S(S+1) = 1$. This normalization ensures unit spin length, enabling a direct correspondence with the classical spins represented on the Bloch sphere. For spin-1, this gives $\hbar_{\text{eff}}= 1/\sqrt{2}$. 

\subsection{Quantum chaos: Spectral analysis and eigenstate delocalization}

\begin{figure}[th]
    \centering
   \includegraphics[width=8.5cm]{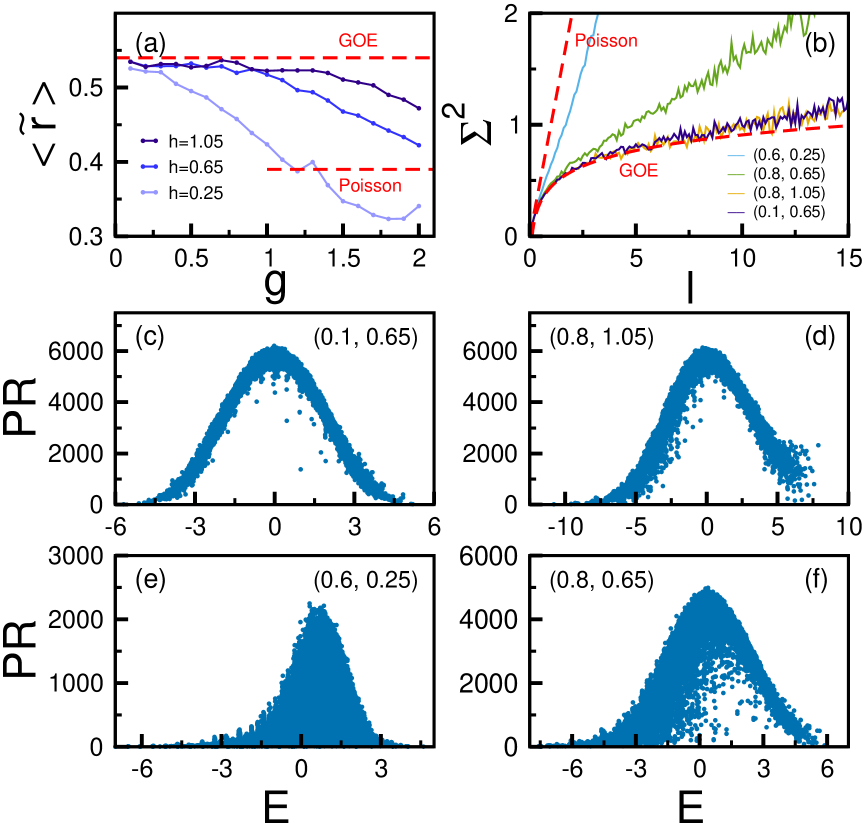}
    \caption{(a)-(b) Analysis of level statistics using (a) ratio of consecutive levels and (b) level number variance, in the odd parity sector. (c)-(f) Participation ratio of the eigenstates in the $z$-basis as a function of the eigenvalues, in both parity sectors. The parameters in (b)-(f) are indicated as $(g,h)$. All panels: $L=9$. 
}
\label{fig:rtilde}
\end{figure}

The onset of quantum chaos is commonly associated with spectral correlations as in random matrix theory~\cite{MehtaBook,Guhr1998}. Since our Hamiltonian matrix is real and symmetric, its spectral properties should be compared with those of the Gaussian orthogonal ensemble (GOE). In Fig.~\ref{fig:rtilde}, we analyze two standard quantum chaos indicators: the ratio of consecutive level spacings~\cite{Oganesyan2007,Atas2013} 
and the level number variance~\cite{MehtaBook,Guhr1998}. 

The ratio of consecutive level spacings~\cite{Oganesyan2007,Atas2013} is defined as 
\begin{equation}
\tilde{r}_{\alpha} = \frac{\min{(s_\alpha,s_{\alpha-1})} }{\max{(s_\alpha,s_{\alpha-1})}  } ,  
\end{equation}
where $s_\alpha=E_{\alpha+1}-E_\alpha$ is the level spacing. For integrable models with Poisson statistics, $\langle \tilde{r} \rangle_P \approx 0.39$, while for chaotic models following the GOE Wigner surmise, $\langle \tilde{r} \rangle_{WD} \approx 0.53$. As shown in Fig.~\ref{fig:rtilde}(a), strong level repulsion emerges when $g, h  \sim J$, but also for relatively small amplitudes of the longitudinal field, where $g<h \sim J$. 

The level number variance gives a more informative picture of the spectrum, capturing both short- and long-range correlations. It is defined as
\begin{equation}
    \Sigma^2(l) = \overline{ N(l,E)^2} - \overline{N(l,E)}^2 ,
\end{equation}
where $N(l,E)$ is the number of unfolded eigenvalues in the interval $[E,E+l]$ and the bar indicates average. For Poisson statistics, for which the eigenvalues are uncorrelated, $\Sigma^2(l)=l$, while for the GOE, the theory predicts $\Sigma^2(l)=2 [\ln(2 \pi l) + \gamma +1 - \pi^2/8]/\pi^2$, where $\gamma$ is the Euler constant. The data for the level number variance in Fig.~\ref{fig:rtilde}(b) confirm the rigidity of the spectrum not only for comparable values of $g$, $h$, and $J$, such as $g=0.8, h=1.05$, but also for small $g$, such as $g=0.1, h=0.65$. 

In addition to analyzing level statistics, we examine the degree of the eigenstate delocalization. For that, we choose as basis the eigenstates of the ``unperturbed'' Hamiltonian  
\begin{equation}
 H_0 =   -g\sum_{j=1}^L S_j^z  -J \sum_{j=1}^{L-1}  S_j^z S_{j+1}^z , 
 \label{eq:MFIM0}
\end{equation}
 denoted by 
\begin{equation}
 \ket{k} \equiv \ket{s_1,...,s_j,...,s_L}_z,
 \label{Eq:basis}
 \end{equation}
 where $-S\leq s_j \leq S$ and $j=1,...,L$. In this basis, the perturbation corresponds to the transverse uniform magnetic field,
$$V=
 - h\sum_{j=1}^L S_j^x ,$$  
which acts locally on each site without introducing inter-particle interactions. This contrasts with more conventional perturbations in the literature, which often involve interaction terms~\cite{Berman2001,Borgonovi2016}. The term $V$ is responsible for the onset of chaos. It couples basis vectors that differ by a single spin flip, that is, by one excitation. 

To quantify the degree of delocalization in the $|k\rangle$ basis, we compute the participation ratio of each eigenstate $|\alpha\rangle$,
\begin{equation}
    \text{PR}_\alpha = \frac{1}{\sum_{k=1}^N |\langle k|\alpha \rangle|^4},
\end{equation}
where $N$ is the dimension of the Hilbert space. The participation ratio measures the effective number of basis states contributing to a given eigenstate, with larger values indicating greater delocalization. 

In the strong chaotic regime, the participation ratio shows small fluctuations as a
function of energy~\cite{ZelevinskyRep1996,Santos2010PRE}. This is indeed what we observe in Figs.~\ref{fig:rtilde}(c)-(d) for the following set of parameters: $(g,h)=(0.1,0.65)$ and $(0.8, 1.05)$, respectively. These results highlight the strong dependence of the eigenstate structure on energy and confirm that highly delocalized, chaotic eigenstates predominantly reside in the central region of the spectrum of many-body quantum systems. For parameters outside the chaotic regime, 
the structure of the eigenstates exhibits strong fluctuations, as seen in Figs.~\ref{fig:rtilde}(e)-(f).

In our subsequent studies of thermalization, we focus on the parameters $g=0.1$ and $h=0.65$, which ensure strong quantum chaos. The rationale for this choice will become clear with the analysis of the classical model in the next section.

\subsection{Eigenstate thermalization}

We extend our analysis of the quantum model by investigating how the chaotic structure of the eigenstates affects the diagonal and off-diagonal matrix elements of few-body observables in the energy eigenbasis. Such studies fall under the framework of the diagonal and off-diagonal eigenstate thermalization hypothesis (ETH) \cite{Alessio2016}. To minimize boundary effects, we consider the $z$- and $x$-components of the magnetization at the central site of the chain. 

\begin{figure}[t]
    \centering
\includegraphics[width=8.cm]{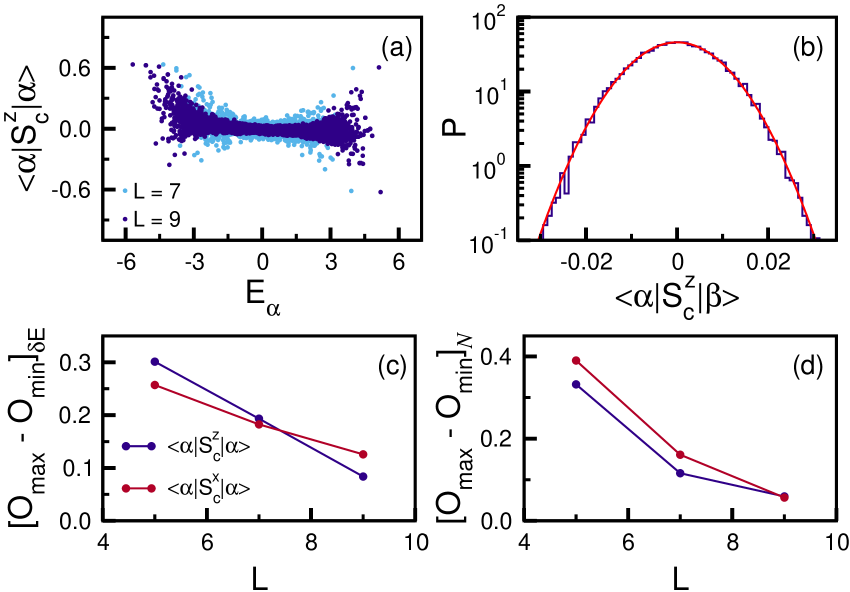}
\caption{Study of the eigenstate thermalization   in the chaotic regime: $g=0.1$ and $h=0.65$. (a) Eigenstate expectation values of $S^z_{c}$ at the center of the chain, $c= \lceil L/2 \rceil$,  as a function of the eigenvalues for $L=7$ (cyan) and $L=9$ (blue).  
(b) Distribution of the off-diagonal elements of $S^z_{c}$ for 200 eigenstates in the middle of the spectrum, $L=9$; the red curve represents a Gaussian distribution. (c)-(d) Extremal fluctuations of  $\langle \alpha |S^z_{c}|\alpha \rangle$ and $\langle \alpha |S^x_{c}|\alpha \rangle$ as a function of system size for (c) a fixed energy window $\delta E = 0.1$ and (d) 40 eigenstates in the middle of the spectrum. Both parity sectors in (a), (c)-(d); odd parity sector in (b).
}
    \label{fig:ETH}
\end{figure}

Figure~\ref{fig:ETH}(a) displays the eigenstate expectation values $\langle \alpha|S^z_{c}|\alpha\rangle$ of the local magnetization in the $z$-direction at the central site. The results mirror the behavior of the participation ratio in Fig.~\ref{fig:rtilde}(c), where the fluctuations are markedly reduced away from the borders of the spectrum. The comparison between system sizes $L=7$ and $L=9$ shows that as the system size increases, the fluctuations of $\langle \alpha|S^z_{c}|\alpha\rangle$ for states close in energy decrease. This trend suggests that, in the thermodynamic limit, $\langle \alpha|S^z_{c}|\alpha\rangle$ for a single eigenstate in the bulk of the energy spectrum should converge to the microcanonical average $(S^z_{c})_{\text{mic}}$, consistent with the expectations of statistical mechanics. For an arbitrary observable $O$, the microcanonical average is obtained as
$$
O_{\text{mic}} = 
\frac{1}{\mathcal{N}_{E,\delta E}}
\sum_{\substack{\alpha \\ |E - E_\alpha| < \delta E}}
 \!\!\!\!\! \langle \alpha|O|\alpha\rangle ,
$$
where $\mathcal{N}_{E,\delta E}$ is the number of energy eigenstates with energy in the window  $\left[E-\delta E, E +\delta E\right]$. This type of analysis belongs to the diagonal formulation of the ETH.

When the system is taken out of equilibrium, its relaxation toward a new equilibrium and the temporal fluctuations around equilibrium are governed by the eigenvalues of the Hamiltonian involved in the dynamics and by the off-diagonal elements of the observable under consideration. This is why when studying thermalization, not only the analysis of the diagonal elements of the observable is important, as performed in Fig.~\ref{fig:ETH}(a), but also the analysis of its off-diagonal elements. 

Figure~\ref{fig:ETH}(b) shows the distribution of the off-diagonal elements of $\langle \beta|S^z_{c}|\alpha\rangle$ obtained for $200$ eigenstates near the middle of the spectrum. The distribution follows a Gaussian shape~\cite{Beugeling2015,LeBlond2019,Santos2020}, reflecting the chaotic nature of the underlying many-body eigenstates, whose components behave as nearly independent Gaussian random variables within the energy shell~\cite{Lydzba2021,Wittmann2022}. 

To test the Gaussianity of the distribution in Fig.~\ref{fig:ETH}(b), we compute the kurtosis, 
\begin{equation}
\kappa = 
\frac{\overline{ (\langle \beta|S^z_{c}|\alpha\rangle - \overline{\langle \beta|S^z_{c}|\alpha\rangle}  )^4 } }{\sigma^4},
\end{equation}
where the bar indicates average over the off-diagonal elements and $\sigma^2$ is the variance. The theoretical value for a Gaussian shape is 3 and we obtain 2.91 for $L=9$. We also evaluate the ratio
\begin{equation}
\Gamma =
\frac{\overline{ \langle \beta|S^z_{c}|\alpha\rangle^2 }}{\overline{ |\langle \beta|S^z_{c}|\alpha\rangle| }^2},
\end{equation}
which for a Gaussian is $\pi/2$. We obtain 1.56 for the system with $L=9$ sites.

The smoothing of the fluctuations in $\langle \alpha|S^z_{c}|\alpha\rangle$ with increasing $L$, shown in Fig.~\ref{fig:ETH}(a), indicates that the infinite-time average of the local magnetization, $\overline{S^z_{c}}=\sum_{\alpha} |C_0^{\alpha}|^2 \langle \alpha|S^z_{c}|\alpha\rangle $, where $C_0^{\alpha} = \langle \alpha|\Psi(0) \rangle$ are the components of the initial state $|\Psi(0) \rangle$, should converge to the microcanonical average $(S^z_{c})_{\text{mic}}$ as the system size increases. But how rapidly does this convergence occur? 

For infinite temperature, the microcanonical value should satisfy $(S^{\mu}_{c})_{\text{mic}} \rightarrow 0$. In Figs.~\ref{fig:ETH}(c)-(d), we show that the extremal fluctuations of the eigenstate expectation values $\langle \alpha| S^{\mu}_{c} |\alpha \rangle$, defined as the difference between the largest and smallest values, $\max S^{\mu}_{c} - \min S^{\mu}_{c}$,  decrease with $L$. In Fig.~\ref{fig:ETH}(c), this range is computed within a narrow energy window of width $\delta E = 0.1$ centered at the middle of the spectrum, yielding $[\max S^{\mu}_{c} - \min S^{\mu}_{c}]_{\delta E}$. In Fig.~\ref{fig:ETH}(d), we instead fix the number of eigenstates to {${\cal N}=40$, with ${\cal N}/2$ states below and ${\cal N}/2$ above the center eigenvalue, and compute $[\max S^{\mu}_{c} - \min S^{\mu}_{c}]_{\cal{N}}$.

Although the fluctuations decrease with increasing system size, performing a systematic scaling analysis of the convergence toward the microcanonical average remains challenging because the Hilbert-space dimension grows exponentially with $L$. Previous studies have suggested a rapid, possibly exponentially, convergence with the system size~\cite{Ikeda2011,Beugeling2014,Rodriguez2024}, while the scaling of the width of the density
of states has been shown to follow an algebraic behavior on $L$ \cite{Rigol2008}. Our present results do not allow for a definitive conclusion regarding the asymptotic scaling behavior. Nevertheless, as we show in the next section, a detailed analysis of the corresponding classical model supports a power-law scaling of the deviations from thermal equilibrium. But before addressing this point, we first turn to the analysis of the occupation number in the quantum model.

\subsection{Occupation number dynamics}
\label{sec:OND}

Motivated by early studies~\cite{Srednicki1994,Horoi1995,Flambaum1997,Borgonovi1998} of thermalization in terms of the occupation number distribution, also discussed in~\cite{Borgonovi2017,Borgonovi2019b}, we adapt this concept to spin models. A spin-1 particle is characterized by three magnetic sublevels $m=-1,0,1$. Thus, for a given many-body state $|\Psi\rangle$, we define the occupation number $\langle n_m^z\rangle$ as the average number of spins (sites $j$) occupying the magnetic sublevel $m$,
\begin{equation}
    \label{eq:ond}
    \langle n_m^z \rangle  = \sum_{j=1}^L| _{j}\langle m|\Psi\rangle|^2 .
\end{equation}
Here $_j\langle m|$ denotes projection of the $j$-th spin onto the local basis state $\ket{k} \equiv \ket{s_1,...,s_j,...,s_L}_z$ with magnetic quantum number $m$.
For example, for a system with two sites in the state 
\begin{eqnarray}
    |\Psi \rangle &=& c_1 \, |-1 -1 \rangle_z + c_2 \,|-1 \quad \,0 \rangle_z + c_3  \,|-1 \quad \,1 \rangle_z \nonumber \\
    && c_4 \, |\quad \, 0 -1 \rangle_z + c_5 \, |\quad \, 0 \quad \, 0 \rangle_z + c_6 \, |\quad \, 0\quad \, 1 \rangle_z \nonumber \\
    && c_7 \, |\quad \, 1 -1 \rangle_z + c_8 \, |\quad \, 1 \quad \, 0 \rangle_z + c_9 \, |\quad \, 1 \quad \, 1 \rangle_z, \nonumber
\end{eqnarray}
the occupation number for the $m=0$ sublevel is
\begin{equation*}
    \langle n_0^z \rangle  = |c_2|^2 +|c_4|^2 + 2 |c_5|^2 + |c_6|^2 + |c_8|^2.
\end{equation*}
By construction, the total population satisfies the normalization condition
$$
\sum_{m=-1}^1 \langle n_m^z \rangle = L.
$$

We begin by examining in Fig.~\ref{fig:dyn}(a) the relaxation of the single-particle occupation numbers $\langle n_m^z (t) \rangle$ toward equilibrium. The system is initialized in a product state corresponding to a quantum quench from an initial Hamiltonian $H_0$ to a final Hamiltonian $H$. Specifically, we choose
\begin{equation}
    \label{eq:in-st}
\ket{\Psi(0)} = \ket{0...0 \ S \ 0...0}_z,
\end{equation}
where the central spin has maximal $z$-magnetization and all other spins have zero $z$-magnetization. This choice corresponds to a localized excitation embedded in an otherwise unpolarized background. 

The energy of this initial state is 
\begin{equation}
    E_0 = \langle \Psi(0) | H| \Psi(0) \rangle = -\hbar_{\text{eff}} g S ,
    \label{Eq:E0}
\end{equation}
which, for the parameters in Fig.~\ref{fig:dyn}(a) ($g=0.1$, $h=0.65$, $L=9$), lies in the middle of the spectrum, $E_0 \sim -0.07$. The energy spread of the initial state, which controls the number of eigenstates participating in the subsequent dynamics, is given by
\begin{equation}
    \sigma^2 = \sum_{k \neq k_0} |\langle k|H|\Psi(0)\rangle|^2 = \frac{h^2}{2(S+1)}[(L-1)S+L] .
    \label{Eq:width}
\end{equation}
yielding $\sigma=1.34$ for the chosen parameters.

The resulting time evolution of the occupation numbers is shown in Fig.~\ref{fig:dyn}(a). According to Eq.~(\ref{eq:in-st}), the system initially exhibits a highly inhomogeneous distribution given by $\langle n_0^z(0)\rangle = 8$ (red curve), $\langle n_1^z(0)\rangle = 1$ (blue curve), and $\langle n_{-1}^z(0)\rangle = 0$ (green curve). As time evolves, the populations relax toward stationary values, marked by the horizontal dashed lines. After saturation, the main panel of Fig.~\ref{fig:dyn}(a) indicates that the three magnetic sublevels become nearly equally populated, consistent with an effective infinite-temperature state in which $\overline{\langle n_m^z\rangle} \approx L/3$. However, the inset reveals a small residual dependence of the steady-state values on $m$. 
This dependence holds for both the infinite-time average and the microcanonical average, indicating that the populations of the magnetic sublevels are not yet perfectly equal. 

\begin{figure}[th]
    \centering
\includegraphics[width=8.cm]{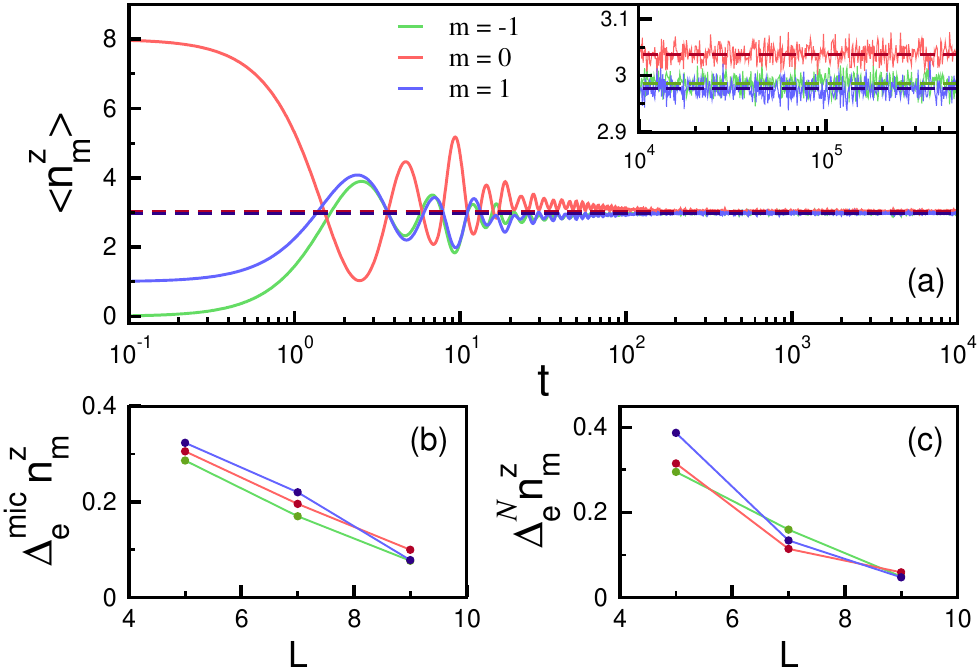}

\caption{
    (a) Main panel and inset: Time evolution of the single particle occupation number for $m=-1,0,1$ (green, red and blue curves,  respectively ); $L=9$.
    Initial state is $\ket{\Psi(0)}= |000010000\rangle$. Horizontal dashed lines are the time averages (diagonal ensemble).
(b)-(c): Normalized extremal fluctuations of the 
eigenstate expectation values of $n_m^z$  as a function of system size for (b) a fixed energy window $\delta E = 0.1$ and (c) $\mathcal{N} = 40$ eigenstates in the middle of the spectrum from both symmetry sectors.
All panels: $g=0.1$, $h=0.65$.
}
    \label{fig:dyn}
\end{figure} 

To verify whether the  discrepancies between the saturating values of the level occupations are due to finite size effects and to quantify how $n_m^z$ approaches $(n_m^z)_{\text{mic}}$ with increasing system size, we analyze how the normalized extremal fluctuations of the eigenstate expectation values $\langle \alpha| n_m^z |\alpha \rangle$ decrease with $L$. In Fig.~\ref{fig:dyn}(b), we evaluate~\cite{Santos2010PREb}
\begin{equation}
\Delta^{\text{mic}}_{e} n_m^z \equiv 
\left| 
\frac{\max n_m^z - \min n_m^z}{(n_m^z)_{\text{mic}}} 
\right| ,
\end{equation}
computed within a fixed energy window of width $\delta E = 0.1$ around the center of the spectrum. In Fig.~\ref{fig:dyn}(c), we fix the number of eigenstates to ${\cal N}=40$ close to the central eigenvalue and calculate 
\begin{equation}
\Delta^{\cal N}_{e} n_m^z \equiv 
\left| 
\frac{\max n_m^z - \min n_m^z}{(n_m^z)_{\cal N}} 
\right| .
\end{equation}
In both cases, Figs.~\ref{fig:dyn}(b)-(c), the normalized extremal fluctuations decrease as the system size increases, signaling a possible convergence toward thermal behavior. However, the restricted range of accessible sizes precludes a reliable finite-size scaling. A detailed scaling analysis is deferred to the next section, where we investigate the classical model.

\section{Classical model: Chaos and ergodicity}
\label{sec:ClChaos}

This section is organized as follows. We begin by characterizing the classical dynamics, computing the Lyapunov exponent to identify the region of strong chaos in parameter space. Having established chaotic dynamics, we next examine ergodicity, since thermalization requires the coexistence of a positive Lyapunov exponent and ergodic exploration of phase space. We analyze how deviations from ergodicity scale with the system size $L$ and find that they decrease algebraically. Building on these classical results, we then compare the classical and quantum time evolution of the local magnetization and demonstrate a robust correspondence between the two. Together, these findings indicate that the power-law scaling observed in the classical model sets a lower bound on the thermalization rate of the quantum system.

For the classical model,  $S_j^x$, $S_j^y$, and $S_j^z$ in Eq.~(\ref{eq:MFIM}) are the three components of the angular momentum of a classical rotor with unit length $|\vec{S}_j|^2 = 1$. The classical equations of motion for the $j$-th spin, 
\begin{equation}
    \frac{d\vec{S}_j}{dt} = 
    \left\{ \vec{S}_j, H \right\},
    \label{eq:clhe}
\end{equation}
are given by
\begin{eqnarray}
\dot{S}_j^x &= & \left[ J\left( S_{j-1}^z+ S_{j+1}^z \right) +g \right]  S_j^y , \nonumber
\\
\dot{S}_j^y &=& 
-\left[ J\left( S_{j-1}^z+ S_{j+1}^z \right) +g \right]  S_j^x +h S_j^z, 
\label{eq:eqm}
\\
\dot{S}_j^z &=&  - h   S_j^y .  \nonumber
\end{eqnarray} 

Taking the second time derivative, we obtain the equations of motion for $L$ driven linear oscillators,
\begin{eqnarray}
\ddot{S}_j^x + \Omega_x^2 S_j^x&= & f_x (\vec{S}_{k-1}, \vec{S}_j, \vec{S}_{j+1} ), \nonumber
\\
\ddot{S}_j^y +\Omega_y^2 S_j^y &=& f_y (\vec{S}_{j-1}, \vec{S}_j, \vec{S}_{j+1} ) , 
\nonumber
\\
\ddot{S}_j^z +\Omega_z^2 S_j^z &=& f_z (\vec{S}_{j-1}, \vec{S}_j, \vec{S}_{j+1} ), 
\label{eq:lin}
\end{eqnarray} 
where the three constant linear frequencies are 
\begin{eqnarray}
\Omega_x^2 &=& g^2 , \nonumber
\\
\Omega_y^2 &=& h^2+g^2 , \nonumber
\\
\Omega_z^2 &=& h^2,
 \label{eq:fre}
\end{eqnarray} 
and the driving spin-dependent forces are
\begin{eqnarray}
f_x &=& h S_j^z\left[ g+J\left( S_{j-1}^z +S_{j+1}^z \right)\right] +hJ S_j^y\left( S_{j-1}^y +S_{j+1}^y \right), \nonumber
\\
f_y &=& hJ S_j^x\left( S_{j-1}^y +S_{j+1}^y \right), \nonumber
\\
f_z &=& h\left[ J\left( S_{j-1}^z +S_{j+1}^z \right)-g\right]S_j^x.
 \label{eq:drfo}
\end{eqnarray} 

To study the classical dynamics, we integrate the equations of motion using a standard fourth-order Runge-Kutta method. Both energy and single spin angular-momentum conservation are verified throughout the evolution, with numerical errors maintained below $10^{-11}$.

\subsection{Classical chaos}
\label{sec:QCI}

To quantify the degree of chaos in the classical model, we analyze the sensitivity of trajectories to initial conditions through the exponential divergence of nearby trajectories in phase space. This divergence is characterized by the maximal Lyapunov exponent, which measures the rate at which initially close trajectories separate over time.

The Lyapunov exponent depends on the parameters $g$, $h$, $J$, and the initial condition. To mimic the quantum initial state described in Sec.~\ref{sec:OND}, we build a classical ensemble in which all spins lie in the $x$-$y$ plane with random orientations, except the central spin, which points along the $z$-direction. 
Random orientations in the $x$-$y$ plane correspond to different initial energies, so this ensemble samples  trajectories from multiple energy shells, each characterized by a maximal Lyapunov exponent $\lambda_{\text{max}} (E)$. 
In Appendix~\ref{AppA}, we show that $\lambda_{\text{max}} (E)$ fluctuates with energy,  which justifies the introduction of an average value $\langle\lambda_{\text{max}} \rangle$. This average quantifies the overall degree of chaoticity associated with the ensemble of initial conditions.

Consistent with the findings of Ref.~\cite{Benet2023}, we confirm that the rate of exponential divergence between trajectories in the full $6L$-dimensional phase space matches the exponential rate of the  separation between trajectories on the 3D Bloch sphere for each single spin. This reduction to an effective single-spin description stems from the presence of $L$ integrals of motion -- namely, the fixed lengths of individual spins -- in additional to the energy conservation. For this reason, our analysis concentrates on the motion of a representative spin, the one located at the central site $c= \lceil L/2 \rceil$.

In Fig.~\ref{fig:lya1}(a), we show $\langle \lambda_{\text{max}}\rangle$ as a function of $h$ for different values of $g$. The system is trivially integrable only when $h=0$. 
We are interested in strong chaos, for which thermalization emerges for any initial condition. 

Our data in Fig.~\ref{fig:lya1}(a) show that the Lyapunov exponent reaches its largest values for $h\sim 0.6$ and $g<1$, which motivates  our choice of parameters $g = 0.1$ and $h = 0.65$, indicated with the vertical arrow. These are the same parameters used in the analysis of the quantum model and are now adopted for the study of the classical model. For larger $g$-values the degree of chaos decreases as visible in Fig.~\ref{fig:lya1}(a).

\begin{figure}[h]
    \centering   \includegraphics[width=8.5cm]{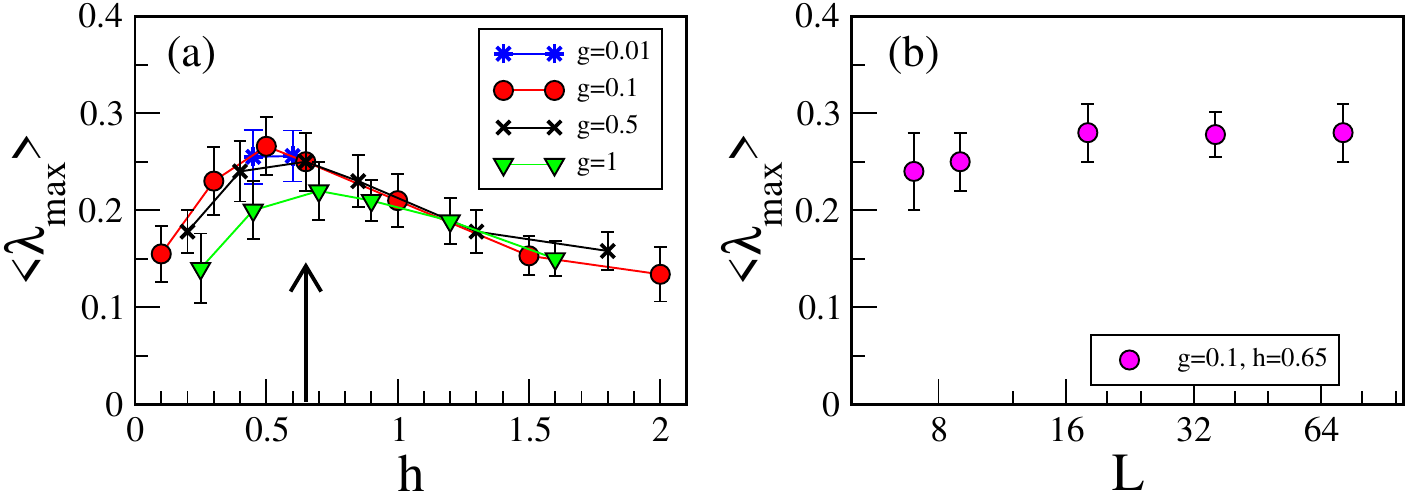}
    \caption{(a) Average maximal Lyapunov  
    exponent  as a function of $h$ for different values of $g$; $L=9$.
    The vertical arrow indicates the values chosen for the studies of the quantum and classical models.
    (b) Average maximal Lyapunov  
    exponent as a function of $L$ for  $g=0.1$ and $h=0.65$.  Averages over $10^3$ initial conditions.
}
    \label{fig:lya1}
\end{figure}

According to Eq.~(\ref{eq:drfo}), chaos is enhanced for $g<J$, since mixing is stronger when the terms depending on both $S^z$ and $S^x$ are larger than the term depending on $S^x$ only. Comparing the curves for $g<J$ in Fig.~\ref{fig:lya1}(a), it is evident that the Lyapunov exponents are weakly dependent on $g$ in this regime.

Figure~\ref{fig:lya1}(b) shows the average maximal Lyapunov exponent $\langle \lambda_{\text{max}}\rangle $ as a function of the system size $L$ for fixed parameters. The results reveal that $\langle \lambda_{\text{max}}\rangle $ is nearly independent of $L$ for sufficiently large system size ($L>10$). The explanation for this behavior follows from  Eq.~(\ref{eq:lin}), which indicates that the motion of each spin can be approximated as a linearly driven oscillator with three different constant frequencies $\Omega_{x,y,z}$ and effective driving forces determined by the motion of neighboring spins. Consequently, the dynamical properties, and in particular the degree of chaos, are very weakly dependent on the system size $L$.


\subsection{Classical ergodicity}
\label{sec:ergCM}


Since the motion of each spin is confined to the surface of its 3D-sphere, ergodicity can be rigorously defined on each sphere. Ergodicity implies that the probability distribution of each Cartesian component of individual spins behaves as the components of random eigenvectors. In random matrix theory, the components of an ergodic eigenstate are distributed according to the following expression~\cite{Benet2023},
  \begin{equation}
      \label{eq:rndm}
      P(S) = \frac{\Gamma(N/2)}{\sqrt{\pi}\Gamma((N-1)/2)}(1-S^2)^{(N-3)/2} ,
  \end{equation}
  where $\Gamma(N)$ denotes the Gamma function (see Appendix~\ref{app:Eq20}).
In our case, $N=3$ and Eq.~(\ref{eq:rndm}) implies a uniform distribution, $P(S)=1/2$, indicating complete ergodicity of motion on the Bloch sphere. In the limit $N \to \infty $, Eq.~(\ref{eq:rndm}) converges to a Gaussian distribution, as expected from the central limit theorem. This reflects the fact that in high-dimensional spaces, the components of random normalized vectors become both statistically independent and normally distributed.

\begin{figure}[t]
    \centering
    \includegraphics[width=8.5cm]{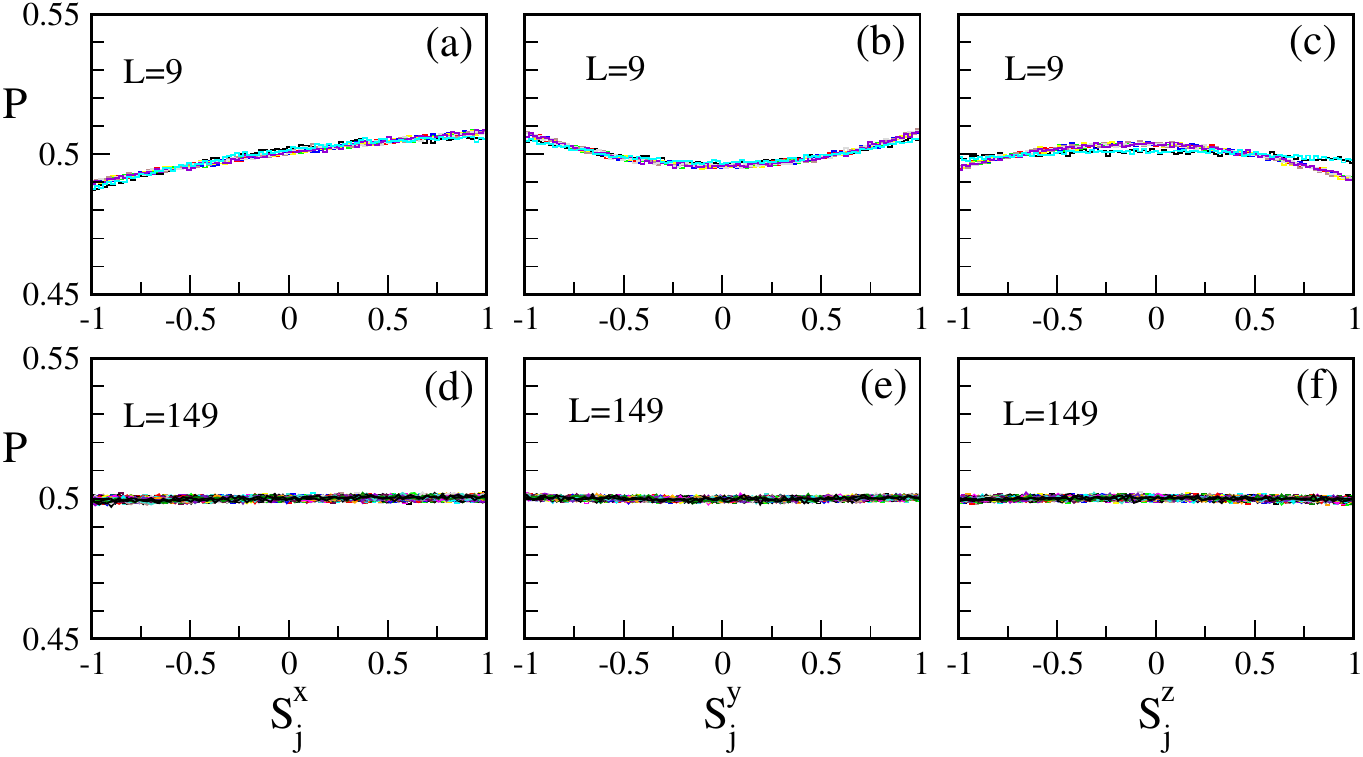}
    \caption{Distribution of single spin components $S_j^{x,y,z}$ for (a)-(c) $L=9$ and (d)-(f) $L=149$. Each color stands for a spin $j=1,...,L$. As initial conditions, we consider $10^4$ trajectories, all with $S_c^z(0) = \hbar_{\text{eff}} S$ and $S_j^z(0) = 0$ for $j\ne c$ (see text). We sample $10^4$ times in increments $dt=1$ after saturation at  $t=10^3$. The number of  histogram bins is $M=10^2$ between the spin values -1 and +1 and the parameters are $g=0.1$ and  $h=0.65$. 
}
    \label{fig:his}
\end{figure} 

To test ergodicity of the classical model on the 3D spin sphere, we analyze many trajectories taken from an ensemble of initial conditions consistent with the quantum initial state considered in Sec.~\ref{sec:OND} (this ensemble is defined and explained in Sec.~\ref{Sec:Class}). Numerical results for the probability distributions of the spin components $S_j^{x,y,z}$ are shown in Fig.~\ref{fig:his} for two system sizes:  $L=9$ [Figs.~\ref{fig:his}(a)-(c)] and $L=149$ [Figs.~\ref{fig:his}(d)-(f)]. 

To obtain the distributions in Fig.~\ref{fig:his}, we generate a total of $N = N_i \times N_t = 10^4 \times 10^4 = 10^8$ data points as follows. We evolve $N_i = 10^4$ trajectories, and for each trajectory, all $L$ spin components are sampled $N_t = 10^4$ times at time intervals $dt=1$ after saturation, for $t > 10^3 $. With $M=100$ histogram bins  between the spin values -1 and +1, each bin contains on average about $10^6$ data points, ensuring robust statistics.

 \begin{figure*}[t]
    \centering
\includegraphics[width=14cm]{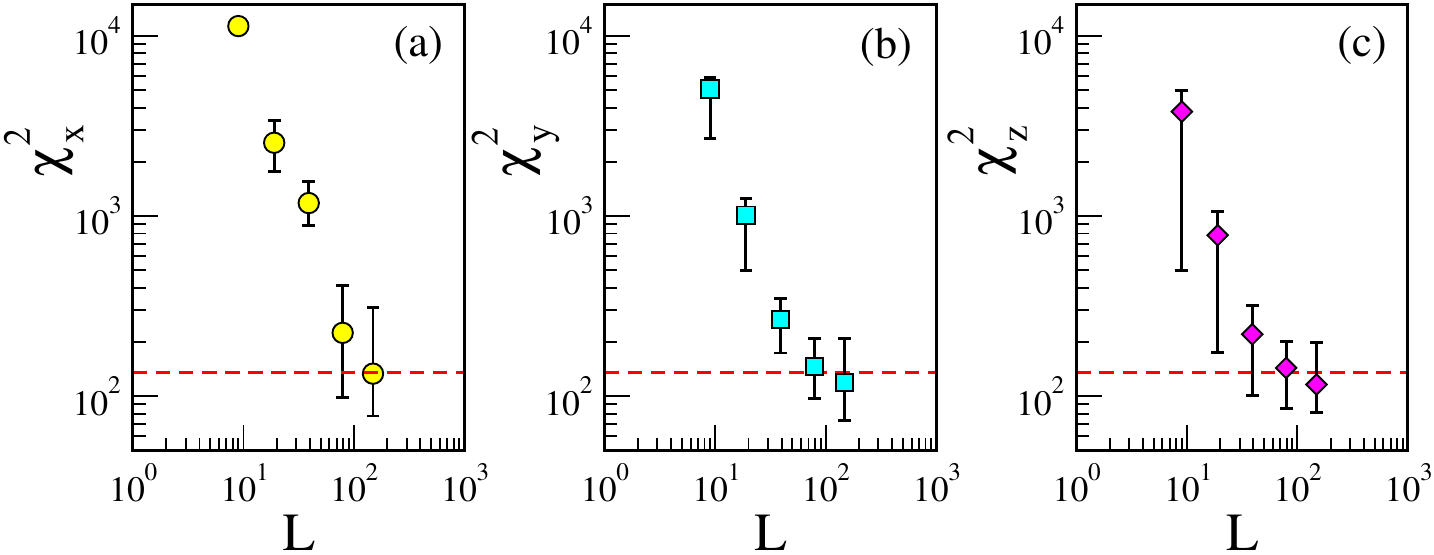}
    \caption{Average $\chi^2_{\mu} = (1/L) \sum_j \chi^2_{j,\mu} $ over all spins as a function of the system size $L$ for the spin components (a) $\mu=x$, (b) $\mu=y$, and (c) $\mu=z$. The
    error bars indicate the interval between the minimal and the maximal values.
    Horizontal dashed line
    represents   the threshold value for ergodicity $(\chi^2)_{99,0.01}$.    We consider  $N_i=10^4$ initial conditions, $N_t=10^4 $ times in increments $dt=1$ for $t>10^3$, and $M=100$  histogram bins between the spin values -1 and +1 .
      The parameters are $  g=0.1$ and $h=0.65$.
}
    \label{fig:xi2}
\end{figure*} 

For $L=9$ in Figs.~\ref{fig:his}(a)-(c), the numerical results for the distributions deviate strongly from the uniform result $P(S)=1/2$ expected for an ergodic filling of the sphere. This indicates incomplete exploration of the available phase space. However, increasing the system size significantly flattens the distributions, as seen in Figs.~\ref{fig:his}(d)-(f) for $L=149$, where the distributions approach the constant value $P(S)=1/2$. This demonstrates that larger systems exhibit increasingly uniform coverage of the sphere, consistent with the onset of ergodic behavior in the thermodynamic limit.

To provide a quantitative test of ergodicity, we compare the numerically obtained distributions of the spin components with the uniform distribution for each component $\mu = x, y, z$ of each spin $j = 1, \ldots, L$. The standard procedure is to perform a Chi-square ($\chi^2$) goodness-of-fit test, defined as
\begin{equation}
    \chi^2_{j, \mu} = \sum_{b=1}^{M} \frac{(n^{\mathrm{obs}}_b - n^{\mathrm{exp}}_b)^2}{n^{\mathrm{exp}}_b},
    \label{eq:chi2}
\end{equation}
where $n^{\mathrm{obs}}_b$ and $n^{\mathrm{exp}}_b$ denote, respectively, the observed and expected numbers of counts in the $b$-th bin. 
The distribution of each spin component is discretized into $M = 100$ bins. 

As explained above, we generate a total of $N = N_i \times N_t = 10^4 \times 10^4 = 10^8$ data points. For a uniform distribution, the expected number of counts in each bin is $n^{\mathrm{exp}}_i = N/M = 10^6$, corresponding to equal probabilities $P(S) = 1/2$ for spin projections, as predicted by random-matrix theory. 

Under the null hypothesis that the data follow a uniform distribution, the statistic $\chi^2$ follows a Chi-square distribution with $df = M - 1 = 99$. At a confidence level $\delta = 0.01$, the critical value is   $(\chi^2)_{df, \delta} =(\chi^2)_{99,0.01} = 135.81$.  If the computed $\chi^2$ is below this threshold, the null hypothesis is accepted, indicating that the distribution is consistent with uniformity and that the system exhibits ergodic behavior.

In Fig.~\ref{fig:xi2}, we plot, for each component $\mu=x,y,z$, the average value $\chi^2_\mu = (1/L) \sum_j \chi^2_{j, \mu}$
computed over all $L$ spins. The error bars represent the minimum and maximum values within the set. As seen in the figure, the hypothesis of a uniform distribution can be rejected for all spins when the system size is small ($L<50$), even at energies corresponding to infinite temperature. As $L$ increases, the average $\chi^2_\mu$ approaches the critical value, indicating an overall trend toward uniformity. Nevertheless, the error bars reveal that even for the largest system considered ($L = 149$), some spins still deviate from the uniform distribution at the 0.01 confidence level. 

A more detailed analysis is presented in Appendix~\ref{app:chi2}. There, we show that the approach to the expected threshold follows an algebraic scaling, with a power-law exponent between 2 and 3 depending on the spin component. The appendix also displays $\chi^2_{j,\mu}$ for each individual spin $j$ at $L = 149$, revealing that the spins whose $\chi^2$ values exceed the critical threshold are predominantly located near the chain boundaries. This behavior reflects the combined influence of finite-size and border effects. 

The coexistence of positive Lyapunov exponents and a power-law convergence to ergodicity signals the onset of classical thermalization in the large-system limit. Since the absence of classical ergodicity implies lack of quantum ergodicity, the same scaling behavior should bound the convergence between infinite-time averages and microcanonical ensemble predictions in the corresponding quantum model, as we discuss next.

\subsection{Quantum-classical correspondence}
\label{Sec:Class}

To analyze the classical dynamics, we follow what was done for the quantum model and consider that the unperturbed part of the Hamiltonian has $h=0$, so that only the longitudinal ($z$-direction) terms are present. The transverse field then acts as a perturbation. In this setting, the dynamics of the $j$-th spin corresponds to a precession about the $z$-axis with a nonlinear frequency 
$$\omega_j^2 = g+J\left[S_{j-1}^z(0)+S_{j+1}(0)\right], $$ 
which is determined by the initial conditions on the $z$-components of the neighboring spins.  This follows directly from the first two  equations in Eq.~(\ref{eq:eqm}). 

To establish a close correspondence with the quantum model, we choose the classical initial conditions as follows.  For all spins,  except the central one, the components are initialized according to
\begin{eqnarray}
\label{eq:in-cl0}
    S_j^z(0) &=&  \epsilon_j , \\\nonumber
    S_j^x(0) &=&  \sqrt{1-S_j^z(0)^2} \ \cos{\theta_j} ,\\\label{eq:init0}
    S_j^y(0) &=& \sqrt{1-S_j^z(0)^2} \ \sin{\theta_j} ,\nonumber
\end{eqnarray}
while for the central spin, $j=c= \lceil L/2 \rceil$,
\begin{eqnarray}
\label{eq:initc}
    S_c^z(0) &=&  \hbar_{\text{eff}} S + \epsilon_c , \\\nonumber
    S_c^x(0) &=&  \sqrt{1-S_c^z(0)^2} \ \cos{\theta_c} ,\\
    S_c^y(0) &=& \sqrt{1-S_c^z(0)^2} \ \sin{\theta_c} .\nonumber
\end{eqnarray}
In the equations above, $\epsilon_{j,c}$ are small random numbers uniformly distributed in the interval $[-10^{-6},10^{-6}]$
and the angles $\theta_{j, c}$ are uniformly distributed in  $[0,2\pi)$. The small perturbations $\epsilon_j$ prevent the dynamics from becoming trapped in periodic orbits, ensuring the generality of the results. The classical observables are obtained by averaging over an ensemble of trajectories generated from these initial conditions. This choice ensures that the classical and quantum averages coincide for the spin components at time $t=0$.

\begin{figure}[h]
    \centering
    \includegraphics[width=8cm]{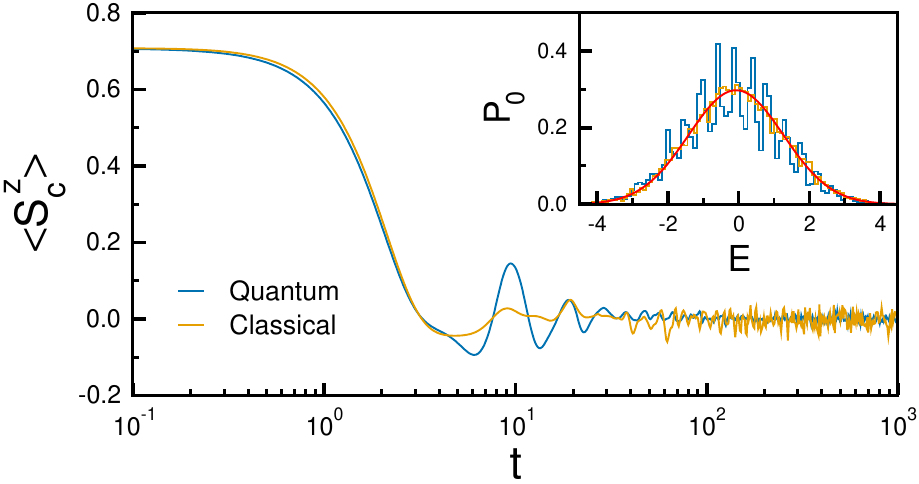}
    \caption{Inset: Energy distribution of the quantum initial state $|000010000\rangle$ (blue line) and of the classical initial state  obtained from an ensemble of classical trajectories with the initial conditions given in Eqs.~(\ref{eq:in-cl0})-(\ref{eq:initc}) (orange curve). The red curve is the Gaussian distribution with first and second moments given by the quantum results in Eqs.~(\ref{Eq:E0})-(\ref{Eq:width}). Main panel: Quantum and classical time evolution of the magnetization in the $z$-direction of the central site $c= \lceil L/2 \rceil$. Parameters: $L=9$, $g=0.1$, $h=0.65$. The number of initial classical trajectories is $10^4$.
}
    \label{fig:ldos}
\end{figure} 

The energy of the quantum initial state is fixed but distributed over the spectrum, as illustrated with the blue curve in the inset of Fig.~\ref{fig:ldos}.  To account for this distribution, the classical initial conditions span a broad range of energies, with each trajectory evolving on a different energy surface. The corresponding classical probability distribution function $P_0(E)$ is shown in orange in the inset. The red curve represents a Gaussian with mean energy $\langle E \rangle = E_0$ [Eq.~(\ref{Eq:E0})] and variance $\sigma^2$ [Eq.~(\ref{Eq:width})]. The close agreement between the quantum and classical distributions, both following the same Gaussian envelope~\cite{Frazier1996,Flambaum2000}, confirms the consistency of the classical ensemble construction. As expected, the quantum distribution exhibits discrete energy bands that gradually smooth out as $S$ increases and the system approaches the classical limit.
 
The main panel of Fig.~\ref{fig:ldos} shows the time evolution of the $z$-component of the magnetization at the central site for both the quantum and the classical models. The close agreement between the two dynamics corroborates the quantum-classical correspondence and indicates that our scaling analysis performed for the classical model in Sec.~\ref{sec:ergCM} should indeed serve as a bound for the quantum system.


\section{Conclusion}
\label{sec:CON}

This work establishes a connection between quantum and classical routes to thermalization in spin systems, using the one-dimensional mixed-field Ising model with spin $S=1$ as a testbed and the occupation number as a central diagnostic. On the quantum side, both the level statistics and the structure of the eigenstates exhibit the signatures of quantum chaos predicted by random matrix theory, indicating the applicability of the eigenstate thermalization hypothesis. However, the exponential growth of the Hilbert space prevents a reliable finite-size scaling analysis. 

To overcome this limitation, we examined the corresponding classical model, where the system size $L$ can be scaled much further. Using a random-matrix-theory-based ergodicity criterion applied to the motion of individual spins on their Bloch spheres, we demonstrated that classical trajectories become increasingly ergodic with system size. Importantly, deviations from ergodicity decay algebraically with $L$, not exponentially. This algebraic scaling provides a quantitative lower bound for the convergence to thermal equilibrium in the quantum model.

By establishing a bridge between the classical and quantum descriptions, we provide a framework to characterize the convergence to thermalization.  This work emphasizes the utility of occupation-number dynamics as a sensitive and physically transparent probe of thermalization in both classical and quantum domains.

\begin{acknowledgements}
F.B.  acknowledges support by the Iniziativa Specifica INFN-DynSysMath. This research was supported in part by NSF grants PHY-1748958 and PHY-2309135 to the Kavli Institute for Theoretical Physics (KITP).    L.F.S and F.M.I.  gratefully acknowledge support from the Simons Center for Geometry and Physics, Stony Brook University at which some of the research for this paper was performed.
\end{acknowledgements}

\appendix
\section{Lyapunov exponent}
\label{AppA}

To assess the degree of chaoticity of the classical model, we examine the divergence of trajectories that start arbitrarily close in phase space and measure the rate of their exponential separation, i.e., the maximal Lyapunov exponent.

\begin{figure}[h]
    \centering
    \includegraphics[width=8cm]{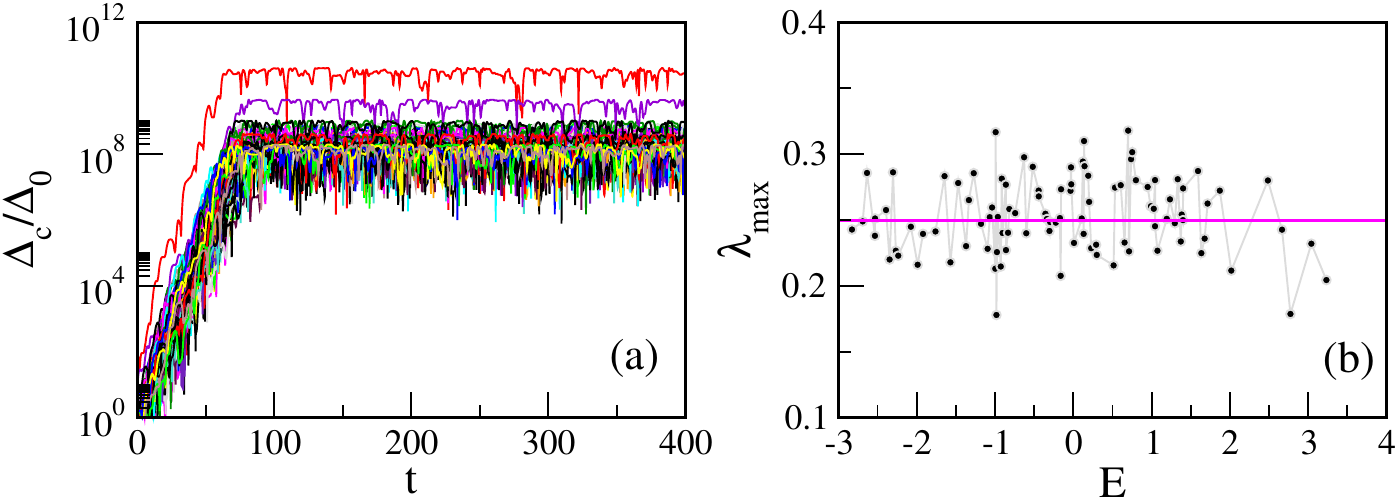}
    \caption{
    (a) Time evolution of the distance between initially nearby trajectories for the central spin $c= \lceil L/2 \rceil$ for $10^2$ trajectories in the ensemble.
    (b) Maximal Lyapunov exponent $\lambda_{\text{max}}$ extracted from the exponential fit of each curve in the range $0<t<70$ in panel (a) as a function of the corresponding initial energy. The horizontal purple line marks the ensemble average $\langle \lambda_{\text{max}} \rangle$.  
}
    \label{fig:lya}
\end{figure} 

To generate two infinitesimally close initial conditions, we introduce a small random tilt ($\epsilon \simeq 10^{-8}$) in all spin components of one initial configuration relative to the other. The separation between the two trajectories in the many-body phase space is quantified by
\begin{equation}
\label{eq:dist}
\Delta^2  = \sum_{j=1}^L \left( S_j^{x\prime} -S_j^x\right)^2 +
\left( S_j^{y\prime} -S_j^y\right)^2+
\left( S_j^{z\prime} -S_j^z\right)^2 ,
\end{equation}
where the prime indicates the trajectory modified by the tilt. One can show that: 
(i) the exponential separation rate in the full $6L$-dimensional phase space coincides with the rate observed on each individual spin sphere; and
(ii) all spins in the chain display the same exponential growth rate.
For this reason, we focus on the central spin only ($c  = \lceil L/2 \rceil$) and monitor
\begin{equation}
\label{eq:distm}
\Delta^2_c (t)   = \sum_{\alpha=x,y,x} 
[ S_c^{\alpha\prime}(t) -S_c^\alpha(t)]^2,
\end{equation}
normalized by the initial displacement
$\Delta_0=\Delta_c(0)$ for an ensemble of initial conditions  satisfying $S_c^z(0) = \hbar_{\text{eff}} S$ and $S_j^z(0)=0$ for $j\ne c$. The time dependence of $\Delta_c(t)$ for all trajectories is shown in Fig.~\ref{fig:lya}(a).

We extract $\lambda_{\text{max}}$ from the exponential growth of $\Delta_c(t)$ via standard exponential fitting of the initial part of the trajectory (before saturation, for $0<t<70$). The resulting values, plotted as a function of the initial energy in Fig.~\ref{fig:lya}(b), display an irregular dependence on energy, which motivates considering the ensemble-averaged Lyapunov exponent $\langle \lambda_{\text{max}} \rangle$ as a representative measure of chaoticity, taking the error bar as one standard deviation in Figs.~\ref{fig:lya1}(a,b).

\section{Derivation of Eq.~(\ref{eq:rndm})}
\label{app:Eq20}

We briefly outline the derivation of Eq.~(\ref{eq:rndm}), which gives the probability distribution of a single component of a random vector uniformly distributed on the unit sphere in \(N\) dimensions. Let \(\mathbf{x}=(x_1,\ldots,x_N)\) be a vector uniformly distributed on the hypersphere defined by
\[
\sum_{i=1}^N x_i^2 = 1 .
\]
We seek the probability density \(P(S)\) of one component, \(S \equiv x_1\).

\begin{figure}[t]
    \centering
\includegraphics[width=8.5cm]{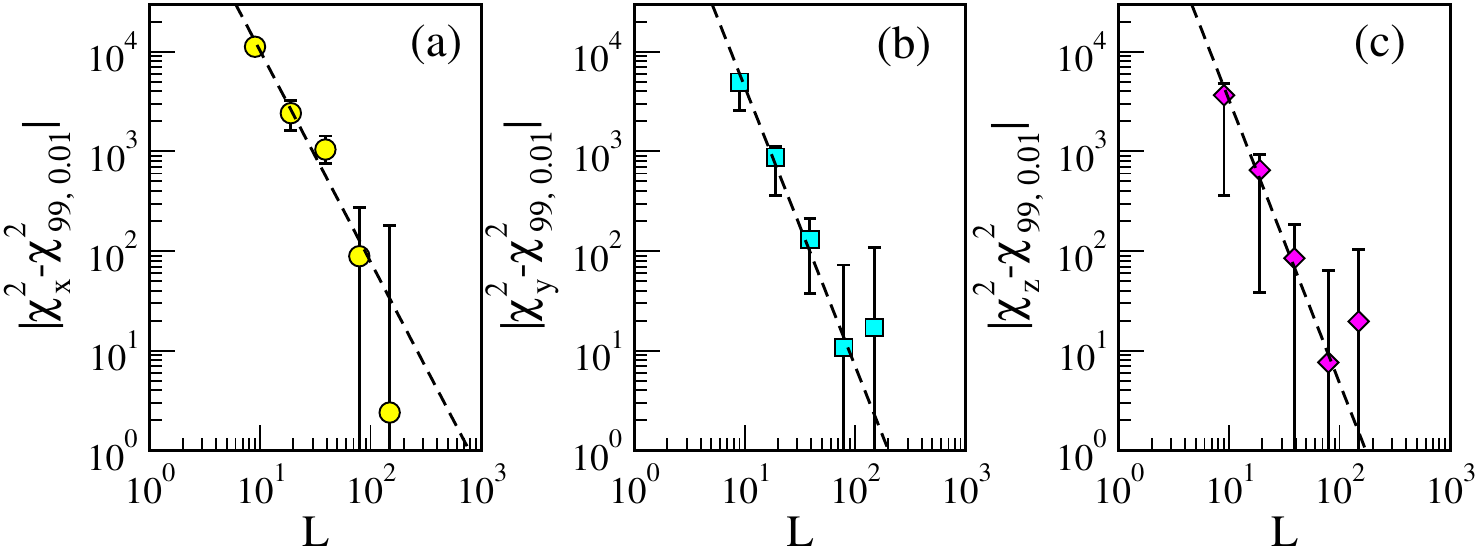}
    \caption{(a)-(c) Deviation 
    from the threshold value as a function of the system size $L$ for the spin components $\mu=x,y,z$. Error bars indicate the range between the minimal and the maximal values.
         We use $N_i=10^4$ initial conditions, $N_t=10^4$ time steps with $dt=1$ in the interval $1000 < t <11000$, and $M=100$  histogram bins for spin values in $[-1,1]$.
    Dashed lines indicate fits of the form $y = A/L^B$ excluding the largest-$L$ point.
    The fitted parameters are:
    (a) $A=1.34 \times 10^6$, $B=2.12$,
    (b) $A=2.86 \times 10^6$, $B=2.80$, and
    (c) $A=2.26 \times 10^6$, $B=2.84$. The parameters are $  g=0.1$ and $h=0.65$.
}
    \label{fig:xi3}
\end{figure} 

Fixing \(x_1 = S\) constrains the remaining components to satisfy
\[
x_2^2 + \cdots + x_N^2 = 1 - S^2,
\]
which, together with the normalization, defines an $(N-2)$-dimensional sphere of radius $r=\sqrt{1-S^2}$. Since the uniform measure on the unit sphere assigns equal weight to all surface elements, the probability density $P(S)$ is proportional to the surface area of this $(N-2)$-sphere. Because the surface area scales as $r^{N-2}$, one obtains
\[
P(S) \propto (1 - S^2)^{(N-3)/2}.
\]
The normalization constant ${\cal C}$ follows from
\[
\int_{-1}^{1} P(S)\, dS = {\cal C} \int_{-1}^{1} (1 - S^2)^{(N-3)/2}\, dS= 1,
\]
which yields Eq.~(\ref{eq:rndm}):
\[
P(S) = \frac{\Gamma\!\left(\frac{N}{2}\right)}{\sqrt{\pi}\,\Gamma\!\left(\frac{N-1}{2}\right)}\,
(1 - S^2)^{(N-3)/2}.
\]

\begin{figure}[t]
    \centering
\includegraphics[width=7.5cm]{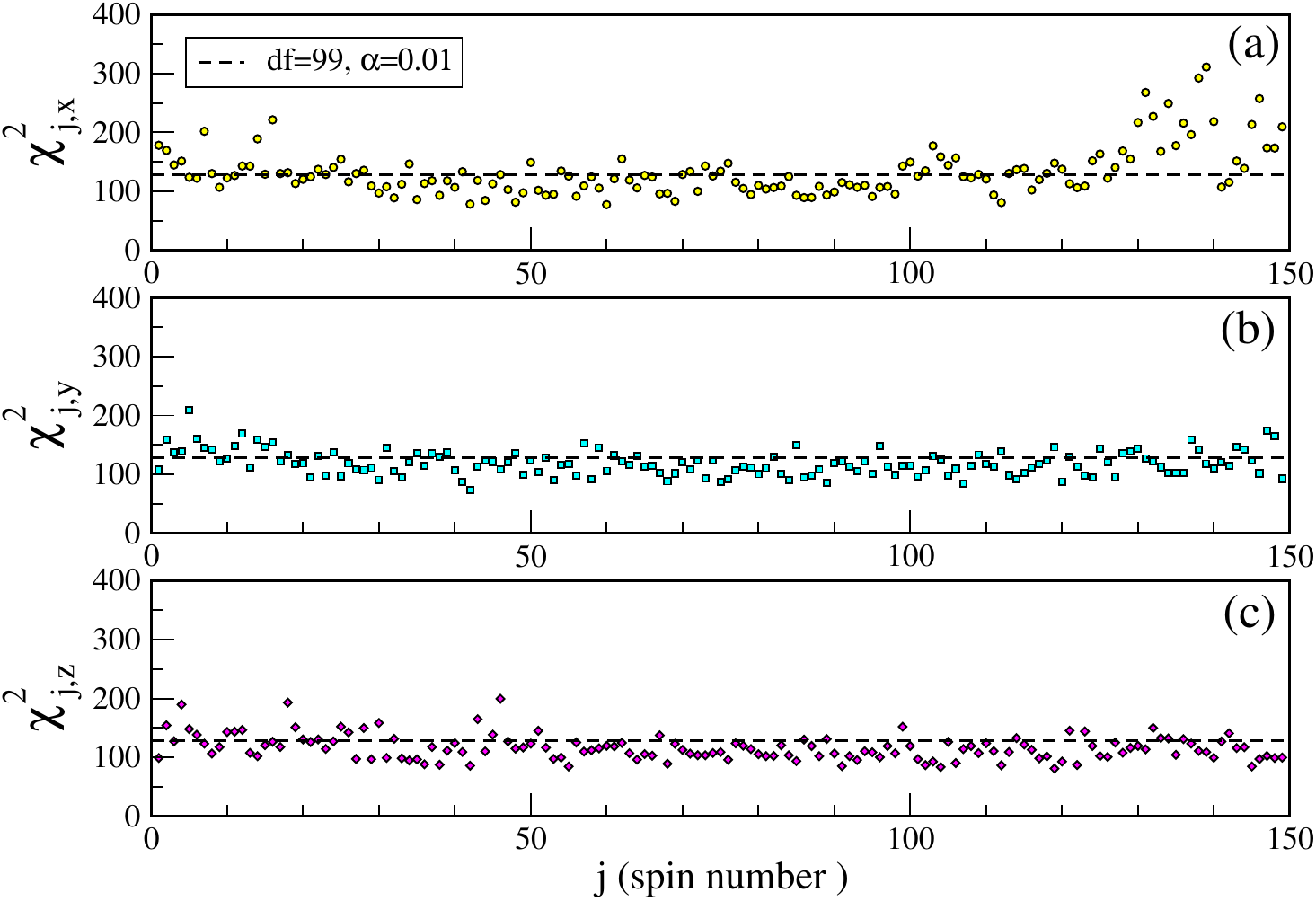}
    \caption{$\chi^2_{j,\mu}$ as a function of the spin position.  Horizontal dashed line is the critical value $\chi^2_{99,0.01}=135.81$.      Parameters:  $L=149$, $g=0.1$, and $h=0.65$.}
    \label{fig:null}
\end{figure} 

For $N=3$, the exponent vanishes and the distribution becomes uniform,
\[
P(S) = \frac{1}{2}, \qquad S \in [-1,1],
\]
reflecting the isotropic distribution of a single spin component on the Bloch sphere.

\section{Algebraic convergence to ergodicity and border effects}
\label{app:chi2}

In Fig.~\ref{fig:xi3}, we plot the deviation
 $| \chi^2_\mu - \chi^2_{99, 0.01}|$ from the critical threshold as a function of system size $L$ for $\mu = x,y,z$. With the exception of the largest system size, the data points fall approximately on a straight line in the log--log scale, indicating algebraic convergence. The dashed curves represent two-parameter fits of the form $A/L^B$ performed on the first four data points, while the last point (largest $L$) is excluded because it saturates the expected value. The resulting exponents are $B = 2.12$, $2.80$, and $2.84$ for the $x$, $y$, and $z$ components, respectively.

It is also informative to examine the site-resolved values $\chi^2_{j,\mu}$ along the chain. In Fig.~\ref{fig:null}, we plot $\chi^2_{j,\mu}$ for all spins and all three components, together with the critical threshold (horizontal dashed line). The deviations above the critical value are concentrated near the chain boundaries, highlighting the finite-size and boundary-condition effects.


%

\end{document}